\documentclass[fleqn,usenatbib]{mnras}

\usepackage{newtxtext,newtxmath}

\usepackage[T1]{fontenc}

\DeclareRobustCommand{\VAN}[3]{#2}
\let\VANthebibliography\thebibliography
\def\thebibliography{\DeclareRobustCommand{\VAN}[3]{##3}\VANthebibliography}

\newcommand{\PRESTO}{\texttt{PRESTO} }

\newcommand{\PULSARMINER}{\texttt{PULSAR\_MINER}}

\newcommand{\h}{^{\rm h}}
\newcommand{\m}{^{\rm m}}
\newcommand{\s}{^{\rm s}}
\usepackage{graphicx}	
\usepackage{amsmath}	

\title[TRAPUM discoveries in NGC 6624]{Four pulsar discoveries in NGC 6624 by TRAPUM using MeerKAT}

\author[Abbate et al.]{\parbox{\textwidth}{
F.~Abbate,$^{1}$\thanks{E-mail: abbate@mpifr-bonn.mpg.de}
A.~Ridolfi,$^{2,1}$
E.~D.~Barr,$^{1}$
S.~Buchner,$^{3}$
M.~Burgay,$^{2}$
D.~J.~Champion,$^{1}$
W.~Chen,$^{1}$
P.~C.~C.~Freire,$^{1}$
T.~Gautam,$^{1}$
J.~M.~Grie{\ss}meier,$^{4,5}$
L.~K\"{u}nkel,$^{6}$ 
M.~Kramer,$^{1,7}$
P.~V.~Padmanabh,$^{8,1}$
A.~Possenti,$^{2}$
S.~Ransom,$^{9}$
M.~Serylak,$^{10,11}$
B.~W.~Stappers,$^{7}$
V.~Venkatraman~Krishnan,$^{1}$
J.~Behrend,$^{1}$
R.~P.~Breton,$^{7}$
L.~Levin,$^{7}$
Y.~Men$^{1}$
}
\\ \\ \\
$^{1}$Max Planck Institut f\"ur Radioastronomie, Auf dem H\"ugel 69 D-53121, Bonn, Germany\\
$^{2}$INAF -- Osservatorio Astronomico di Cagliari, Via della Scienza 5, I-09047 Selargius (CA), Italy\\
$^{3}$ South African Radio Astronomy Observatory (SARAO), 2 Fir Street, Black River Park, Observatory, Cape Town, 7925, South Africa\\
$^{4}$LPC2E - Universit\'{e} d'Orl\'{e}ans /  CNRS, 45071 Orl\'{e}ans cedex 2, France\\
$^{5}$Station de Radioastronomie de Nan\c{c}ay, Observatoire de Paris, PSL Research University, CNRS, Univ. Orl\'{e}ans, OSUC, 18330 Nan\c{c}ay, France\\
$^{6}$Fakult\"{a}t f\"{u}r Physik, Universit\"{a}t Bielefeld, Postfach 100131, D-33501 Bielefeld, Germany\\
$^{7}$Jodrell Bank Centre for Astrophysics, Department of Physics and Astronomy, The University of Manchester, Manchester M13 9PL, UK\\
$^{8}$Max-Planck-Institut f\"{u}r Gravitationsphysik (Albert-Einstein-Institut), D-30167 Hannover, Germany\\
$^{9}$National Radio Astronomy Observatory, 520 Edgemont Rd., Charlottesville, VA 22903, USA\\
$^{10}$SKA Observatory, Jodrell Bank, Lower Withington, Macclesfield, SK11 9FT, United Kingdom\\
$^{11}$Department of Physics and Electronics, Rhodes University, PO Box 94, Grahamstown 6140, South Africa\\
}

\date{Accepted XXX. Received YYY; in original form ZZZ}

\pubyear{2022}

\begin{document}
\label{firstpage}
\pagerange{\pageref{firstpage}--\pageref{lastpage}}
\maketitle

\begin{abstract}
We report 4 new pulsars discovered in the core-collapsed globular cluster (GC) NGC 6624 by the TRAPUM Large Survey Project with the MeerKAT telescope. All of the new pulsars found are isolated. PSR J1823$-$3021I and PSR J1823$-$3021K are millisecond pulsars with period of respectively 4.319 ms and 2.768 ms. PSR J1823$-$3021J is mildly recycled with a period of 20.899 ms, and PSR J1823$-$3022 is a long period pulsar with a period of 2.497 s. The pulsars J1823$-$3021I, J1823$-$3021J, and J1823$-$3021K have position and dispersion measure (DM) compatible with being members of the GC and are therefore associated with NGC 6624. Pulsar J1823$-$3022 is the only pulsar bright enough to be re-detected in archival observations of the cluster. This allowed the determination of a timing solution that spans over two decades. It is not possible at the moment to claim the association of pulsar J1823$-$3022 with the GC given the long period and large offset in position ($\sim 3$ arcminutes) and DM (with a fractional difference of 11 percent compared the average of the pulsars in NGC 6624). 
The discoveries made use of the beamforming capability of the TRAPUM backend to generate multiple beams in the same field of view which allows sensitive searches to be performed over a few half-light radii from the cluster center and can simultaneously localise the discoveries. The discoveries reflect the properties expected for pulsars in core-collapsed GCs.

\end{abstract}

\begin{keywords}
Star:neutron - Pulsars:general - globular clusters:individual: NGC 6624
\end{keywords}



\section{Introduction}

NGC 6624 is a globular cluster (GC) located in the bulge close to the Galactic center (Galactic coordinates $l=2^{\circ}.78$, $b=-7^{\circ}.91$) at a distance from the Sun of $8.0 \pm 0.1$ kpc \citep{Baumgardt2021} and a Galactocentric radius of 1.2 kpc \footnote{A list of the structural parameters of GCs is available at the website: \url{ https://people.smp.uq.edu.au/HolgerBaumgardt/globular/parameter.html}}.
The stellar distribution in the central regions of the GC is not flat but shows a "cusp" that is typical of systems that have undergone core collapse \citep{Sosin1995,Noyola2006}. 

{ This GC is known to contain eight pulsars \citep{Biggs1994,Lynch2012,Ridolfi2021} of which six are millisecond pulsars (MSPs) and two long-period pulsars. Only two pulsars J1823$-$3021F and J1823$-$3021G are known to be part of binary systems. J1823$-$3021G is a binary pulsar in an eccentric orbit typical of an exchange encounter that has the potential of being one of the most massive known pulsars \citep{Ridolfi2021}.}
Two pulsars have long spin periods, J1823$-$3021B has a period of 378.59 ms and J1823$-$3021C has a period of 405.93 ms. Both have short characteristic ages, respectively 190 and 30 Myr, which is at odds with the regular pulsar formation scenario. The pulsars should have formed recently but the lack of star formation { in this globular cluster} in the last few Gyrs implies that no massive stars were around to form them. 
{ This could suggest that these pulsars were old, unrecycled NSs that were, thanks to exchange encounters, incorporated in X-ray binaries, where they started being spun up by accretion of matter from their new companions. Because of the high stellar densities in these clusters, these X-ray binaries were disrupted before the pulsar's magnetic field decreased to the typical values observed in ``fully recycled'' MSPs} \citep{Verbunt2014}. Alternative formation scenarios have been put forward to explain the presence of such pulsars like accretion-induced collapse, direct collisions with a main sequence star \citep{Lyne1996}, or electron capture supernova of an OMgNe white dwarf \citep{Boyles2011}. 

The isolated MSP J1823$-$3021A also has a very high spin period derivative, { and also a small characteristic age of 25 Myr}. It was suggested that the period derivative is heavily influenced by the gravitational potential of the GC and that it could be a sign of the presence of an intermediate mass black hole in the centre of NGC 6624 \citep{Perera2017,Perera2017b}. However, the pulsar also shows remarkably strong gamma-ray flux implying that a significant fraction of the spin down must be intrinsic and the magnetic field has to be considerably higher than that of the average MSP \citep{Freire2011}. A higher number of pulsars close to the cluster centre would allow to test whether an intermediate mass black hole is really present.

The properties of the pulsars in NGC 6624 are quite different from the typical GC but are similar to those GCs that have undergone core-collapse. 
\cite{Verbunt2014} have shown that this difference can be quantified by the encounter rate for a single binary in the GC which represents the { likelihood of any particular binary system} to undergo dynamical encounters; this is $\sim 4$ times higher for core collapsed clusters than non collapsed clusters. GCs with high values of this parameter usually show a larger fraction of isolated pulsars, pulsars with longer spin periods, pulsars distributed over large radii outside of the core, and binaries produced in exchange encounters.

The high value of the encounter rate of this GC (the ninth highest of all the Galactic GCs and $\sim$ 1.1 times the value of 47 Tucanae, \citealt{Bahramian2013}) suggests that a larger population of pulsars is still undiscovered in this GC, while the high encounter rate per binary (second highest within all GCs with 3 or more known pulsars and $\sim 6$ times the value of 47 Tucanae, \citealt{Verbunt2014}) suggests the new discoveries might have gone through dynamical interactions like exchange encounters that could form exotic objects such as eccentric double neutron star systems. 
For these reasons, NGC 6624 was considered as a high priority target in the pulsar searches in GCs with the MeerKAT radio telescope in South Africa \citep{Jonas2016,Camilo2018}. Two of the Large Survey Projects (LSP) that are observing with the telescope, MeerTime\footnote{\url{http://www.meertime.org}} \citep{Bailes2016, Bailes2020} and TRansients And PUlsars with Meerkat (TRAPUM)\footnote{\url{http://www.trapum.org}} \citep{Stappers2016} have the search and timing of pulsars in GCs as one of the science goals. { Together the two projects have discovered more than 40 new pulsars in GCs\footnote{An up-to-date list of the new pulsars discovered can be seen at: \url{http://www.trapum.org/discoveries/}}.}

Given the scientific overlap, the two projects have decided to share resources and observing time. The collaboration has already resulted in two publications regarding NGC 6624: a detailed study of the giant pulses from J1823$-$3021A \citep{Abbate2020} and the discovery of two new MSPs \citep{Ridolfi2021} in this cluster. 
{ The massive eccentric binary J1823$-$3021G discovered in the latter work is an example of the interesting new pulsars still to be discovered.}
These results were obtained using observations taken with the Pulsar Timing User Supplied Equipment (PTUSE) machines made available by MeerTIME with data from only the central 44 antennas of the 64 available. 
This number was chosen so that the maximum baseline would be below 1 km and the tied array beam would be large enough to cover the central regions of the cluster. The observations were carried out at L-band (856-1712 MHz) and the tied array beam had a radius of $\sim 0.5 $ arcmin.

In this paper we present the observations and discoveries made using the Filterbanking Beamformer User Supplied Equipment (FBFUSE) and Accelerated Pulsar Search User Supplied Equipment (APSUSE) backend provided by TRAPUM. Together these machines are able to synthesize and record of order 500 coherent beams for use in pulsar and fast transient searching. Using all of the available antennas it is possible to cover, with a large number of small tied-array beams, an area as large as $1-4$ arcmin in radius depending on the configuration \citep{Chen2021}. This is enough to extend up to more than twice the reported half-light radius of the GC of $\sim 1$ arcmin \citep{Baumgardt2018}. 
The sensitivity is $\sim 1.5$ times higher when compared with the previous PTUSE observations in the center of the GC and it gets increasingly higher the further away we look. The area that can be searched is also larger allowing for further discoveries in this GC.

\begin{figure*}
\centering
	\includegraphics[width=0.6\textwidth]{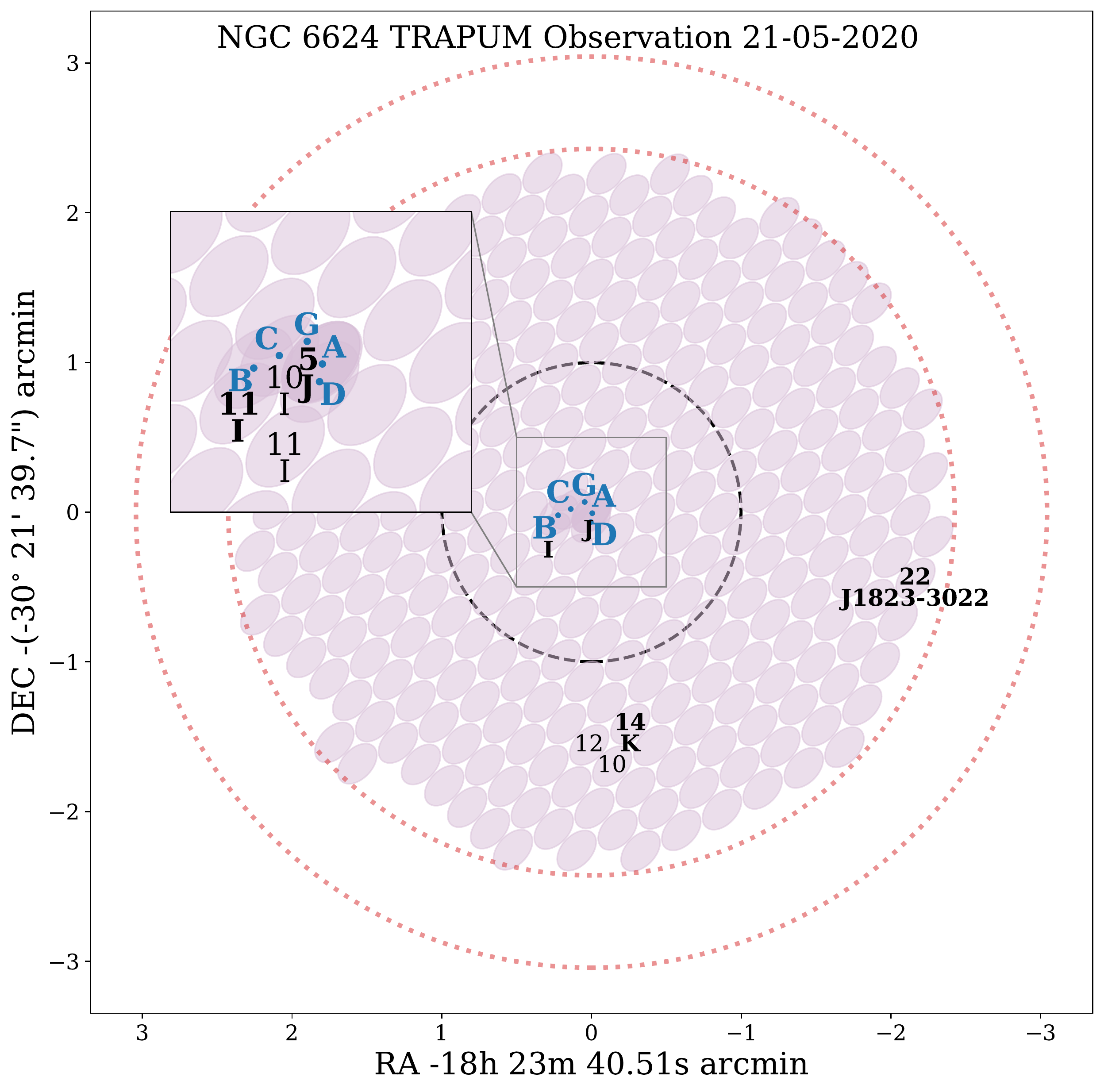}
	\,
	\includegraphics[width=0.6\textwidth]{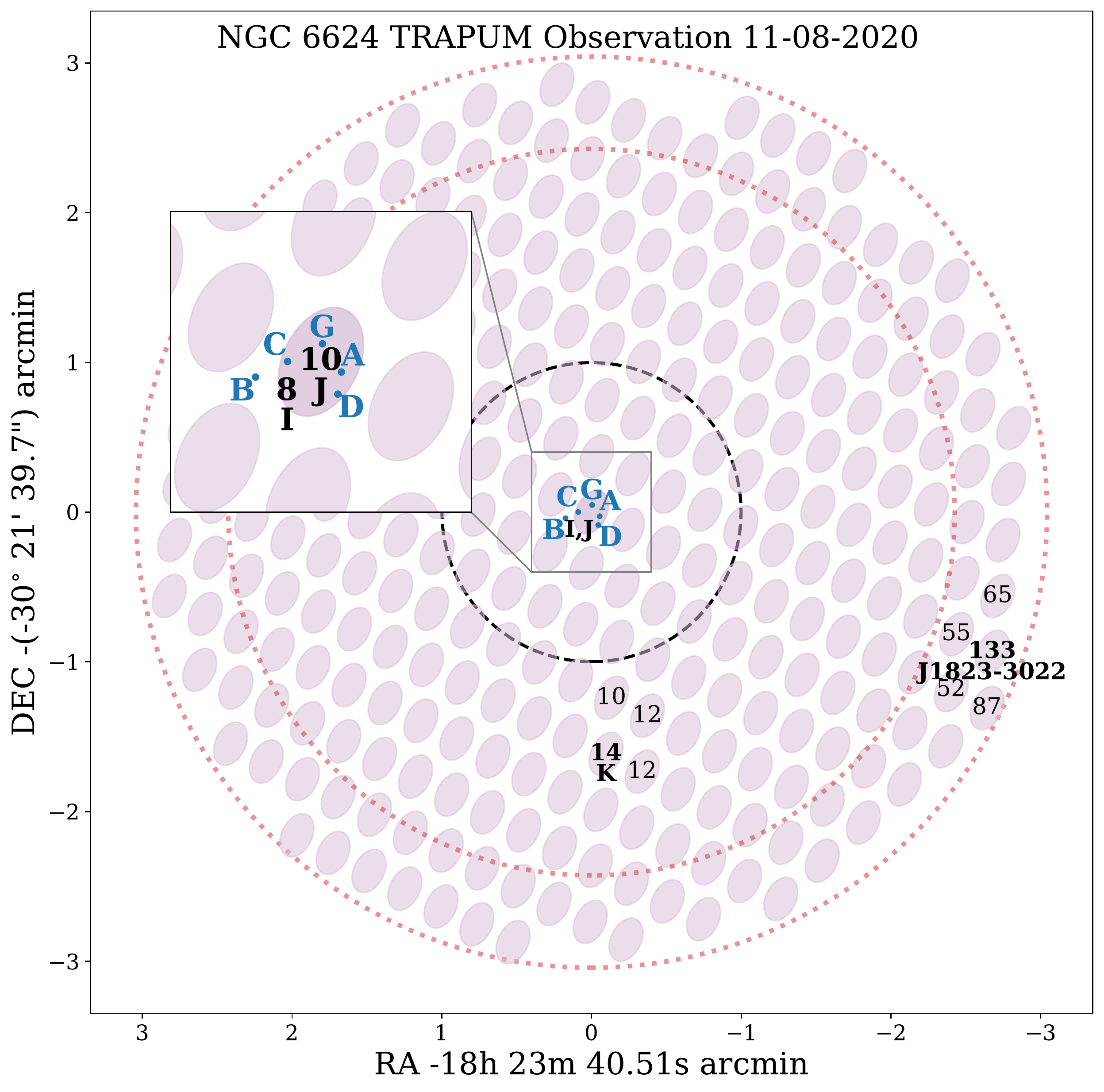}
  	\caption{Tiling pattern of the beams of the TRAPUM observations of NGC 6624 performed on the 21st of May 2020 (top) and on the 11th of August 2020 (bottom). The light pink ellipses show the single beams up to 80 percent of the boresight power. The blue dots are the pulsars with precisely known positions. The black dashed circle shows the half-light radius of the cluster, $\sim 1$ arcminute \citep{Baumgardt2018}. The red dotted circles are the maximum extent of the size of the tiling in the two observations plotted for comparison. The value of the S/N of the detection of the new pulsars are reported in each beam where the pulsars are seen with the highest value reported in bold. All of the beams shown have been searched for new pulsars.}
  	\label{fig:tiling_patterns}
\end{figure*}

\section{TRAPUM observations}

The observations performed with the FBFUSE and APSUSE machines of TRAPUM that have been searched for new pulsars were performed using 60 antennas in L-band (856-1712 MHz) on the 21st of May 2020 and the 11th of August 2020. In both observations we synthesized 288 coherent beams. The beam shape depends on the exact configuration of the antennas and position in the sky, we can estimate its size with an elliptical fit. In both observations the semi-major axis of the elliptical fit at 50 percent of the power is 17 arcsec at the central frequency of 1284 MHz and semi-minor axis is 10 arcsec. In the first observation, the average overlap fraction between neighbouring beams at the central frequency was $\sim 0.8$ resulting in a total area covered of $\sim 2.6$ arcmin in radius from the center. For the second observation the overlap was set at $\sim 0.7$ and the area covered was increased to $\sim 3.1$ arcmin in radius. The tiling pattern of the two observations was determined using the software \texttt{Mosaic} \footnote{\url{https://github.com/wchenastro/Mosaic}} \citep{Chen2021} and is shown in Figure \ref{fig:tiling_patterns}. In both cases, the observation length was 4 hours. The first observation was already described in \cite{Ridolfi2021} and was used to localise the known pulsars.

The total intensity data from each coherent beam was recorded in filterbank mode with a time resolution of  76 $\mu$s and a bandwidth of 856 MHz, centred around 1284 MHz and divided into 4096 channels. 
After the observation was concluded, the data were incoherently de-dispersed at the dispersion measure (DM) of 86.88 pc cm$^{-3}$, corresponding to the DM of J1823$-$3021A located near the center of the cluster { as measured by MeerKAT observations in the same band \citep{Abbate2020}}. The Inter-Quartile Range Mitigation algorithm\footnote{\url{https://github.com/v-morello/iqrm}} \citep{Morello2021} was applied to filter out the brightest Radio Frequency Interference (RFI) { and the channels were averaged by a factor of 16 resulting in 256 channels across the full bandwidth} to reduce disk usage. 

Most of the known pulsars in NGC 6624 have DMs between 86.2 and 87.0 pc cm$^{-3}$ with the exception of J1823$-$3021E that has a DM of 91.4 pc cm$^{-3}$. We decided to search for pulsars between a DM of 80 and 95 pc cm$^{-3}$ with a DM step size of 0.05 pc cm$^{-3}$ to allow for a greater spread in DM in the potential candidates. Given the reduced number of channels, the pulse of a potential pulsar will be affected by smearing if the DM does not match the value applied during de-dispersion of 86.88 pc cm$^{-3}$. The smearing caused by the DM in the worst affected channel goes from 0.3 ms for a DM of 80 pc cm$^{-3}$ to 0.4 ms for a DM of 95 pc cm$^{-3}$. This effect will reduce the sensitivity of our search for rapidly spinning pulsars if the DM is far from the value chosen before reducing the number of channels.

The searching procedure was carried out using \PULSARMINER \footnote{\url{https://github.com/alex88ridolfi/PULSAR\_MINER}} based around the \PRESTO pulsar searching package \citep{Ransom2002} and described in detail in \cite{Ridolfi2021}. The beams were searched for both isolated and binary pulsars by performing a search in Fourier space. In the case of binary pulsars, the acceleration along the line of sight, $a_l$, shifts the observed spin frequency of a potential pulsar over the length of the observation, $ t_{\rm obs}$, by a number of  $z =t_{\rm obs}^2 a_l / cP$ Fourier bins, where $c$ is the speed of light and $P$ is the period of the potential pulsar. The search is performed up to a maximum of $z_{\rm max}$ of 200. This type of search is effective as long as the acceleration remains constant over the observation which is a reasonable assumption if $ t_{\rm obs} \lesssim 0.1 P_{\rm b}$, where $P_{\rm b}$ is the orbital period \citep{Ransom2003}. In order to search for binary pulsars with short orbital periods, we split the observation in to segments of 60 minutes and 20 minutes and repeated the search in each segment. This strategy allowed us to be sensitive to pulsars in orbits longer than $\sim 200$ minutes. 

Of all the candidates that were found in the search only the ones with a Fourier significance of more than $4\sigma$ in adjacent DM trials were kept. After removing harmonically related candidates and periodic RFI, the remaining candidates were visually inspected. In order to confirm that the candidates found are real pulsars, we checked if the same candidate appears in neighbouring beams of the same observation or in beams pointing in the same direction in both observations. Additionally, for the most interesting candidates, we folded the corresponding beam of the other observation at the period and DM of the candidate for further confirmation. After a candidate has been confirmed, we derived an approximate ephemeris based on the detection, folded the observation using the \texttt{DSPSR}\footnote{\url{http://dspsr.sourceforge.net}} pulsar package \citep{vanStraten2011} and run the \texttt{pdmp} routine of \texttt{PSRCHIVE}\footnote{\url{http://psrchive.sourceforge.net}} \citep{Hotan2004,vanStraten2012} to improve the quality of the detection.  { To obtain the best value of DM we first averaged the frequency channels of the best detection from 256 to 8 and extracted times-of-arrival (ToAs) at each channel with the \texttt{pat} routine of \texttt{PSRCHIVE} using a single template. We used \texttt{TEMPO2}\footnote{\url{https://bitbucket.org/psrsoft/tempo2/}} \citep{Hobbs2006} to fit these ToAs leaving the DM as the only free parameter. }

Finally, the flux density of the pulsars was measured using the radiometer equation \citep{Handbook2012}:

\begin{equation}
    S= {\rm (S/N)_{\rm best}} \frac{T_{\rm sys}}{G\sqrt{n_p t_{\rm obs} \Delta f}}\sqrt{\frac{W}{P-W}},
\end{equation}
where (S/N)$_{\rm best}$ is the signal to noise ratio (S/N) of the best detection of the pulsar, $T_{\rm sys}$ is the system temperature of the telescope taken to be 26~K for the observations\footnote{\url{https://skaafrica.atlassian.net/rest/servicedesk/knowledgebase/latest/articles/view/277315585}}, $G=(N_{\rm ant}/64)\times 2.8$ K~Jy$^{-1}$ is the gain of the telescope and depends on $N_{\rm ant}$ the number of antennas used in the observation, $n_p=2$ is the number of polarisations summed, $t_{\rm obs}$ is the length of the observation, $\Delta f$ is the effective bandwidth taken to be 700 MHz after the RFI cleaning { and $P$ is the period of the pulsar. $W$ is the equivalent width of the pulse defined as the width of a top-hat function with the same height and same total area as the pulsar profile.} 

\section{Discoveries}

\begin{figure*}
\centering
	\includegraphics[width=0.4\textwidth]{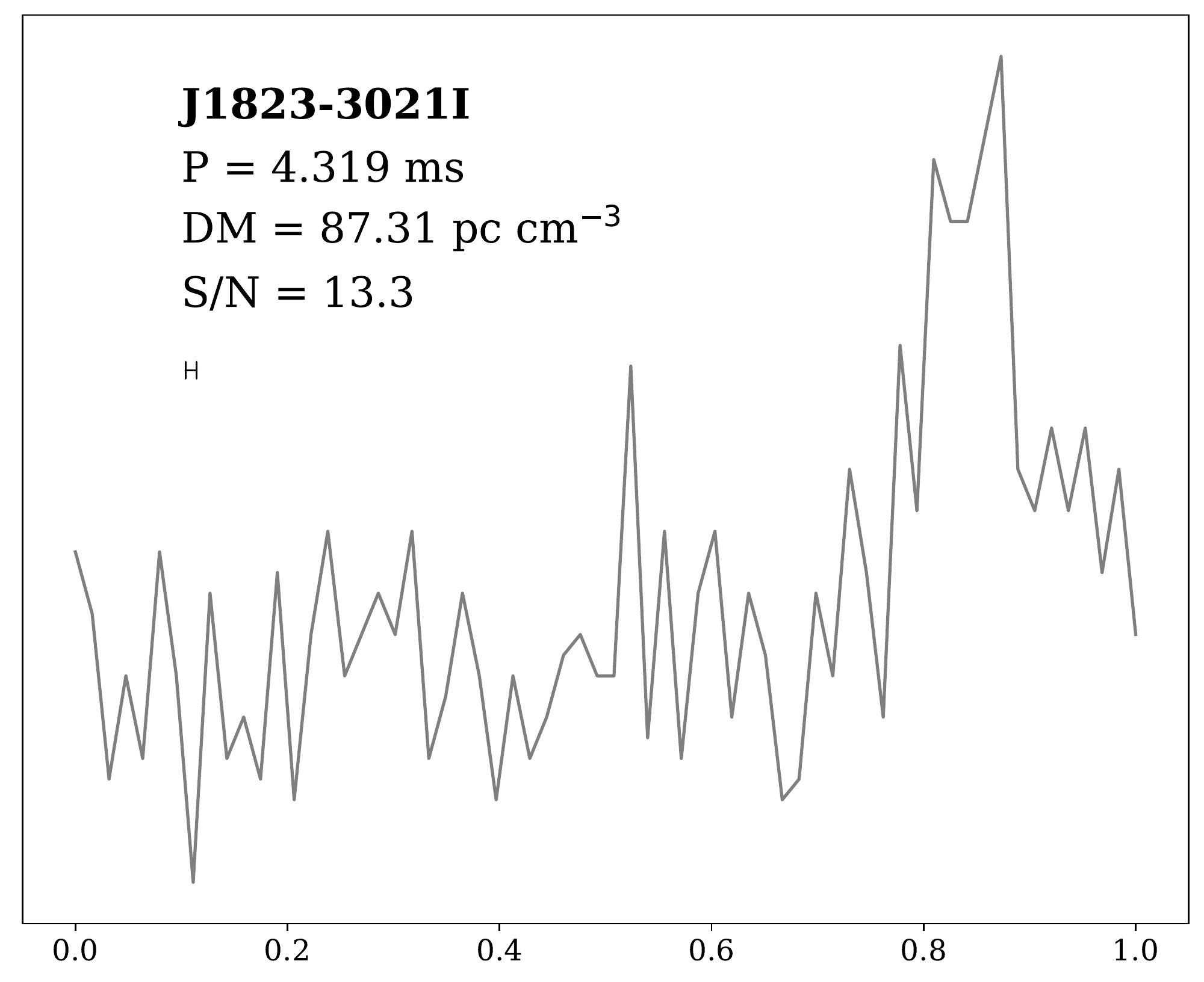}
	\,
	\includegraphics[width=0.4\textwidth]{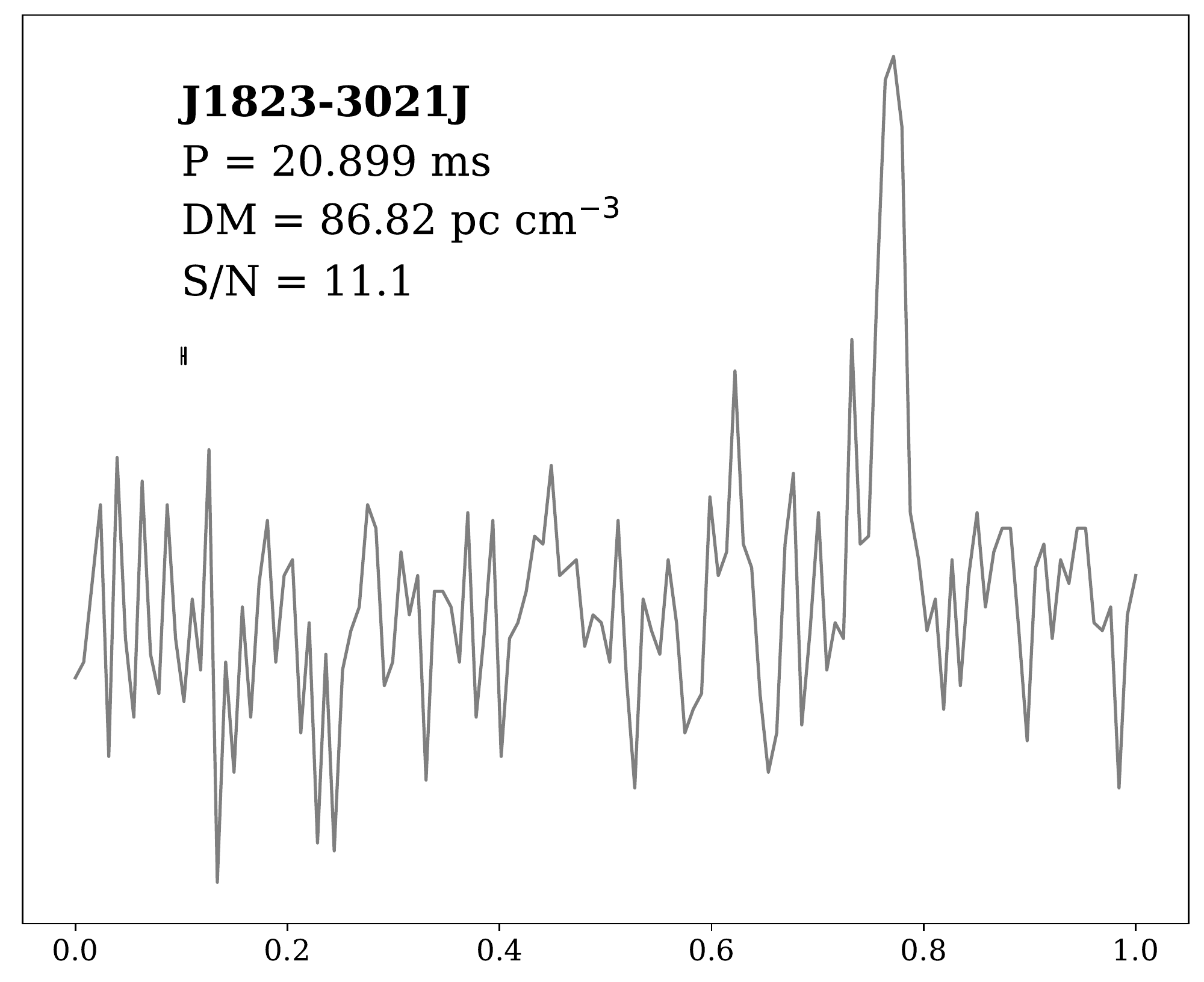}
	\,
	\includegraphics[width=0.4\textwidth]{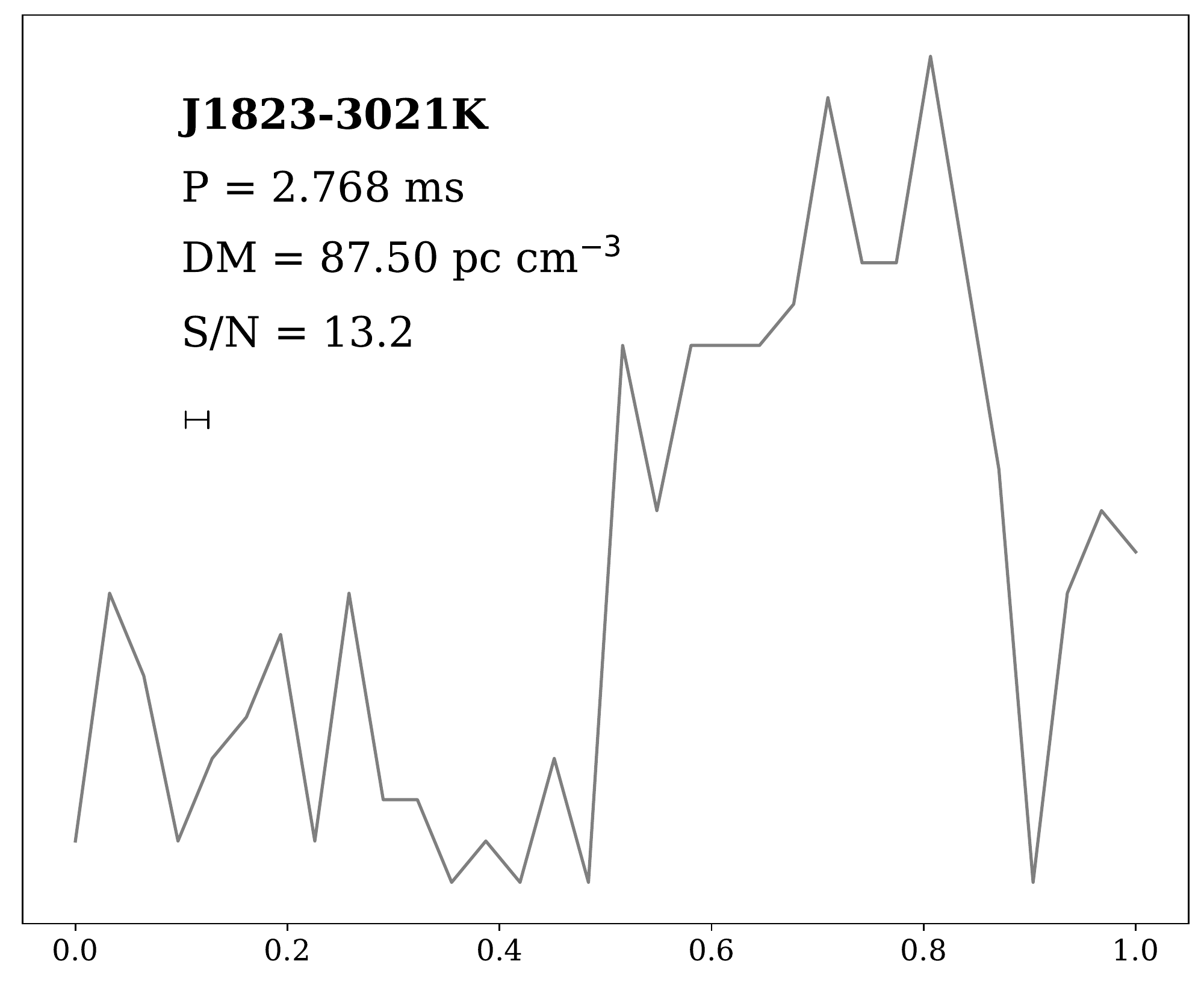}	
	\,
	\includegraphics[width=0.4\textwidth]{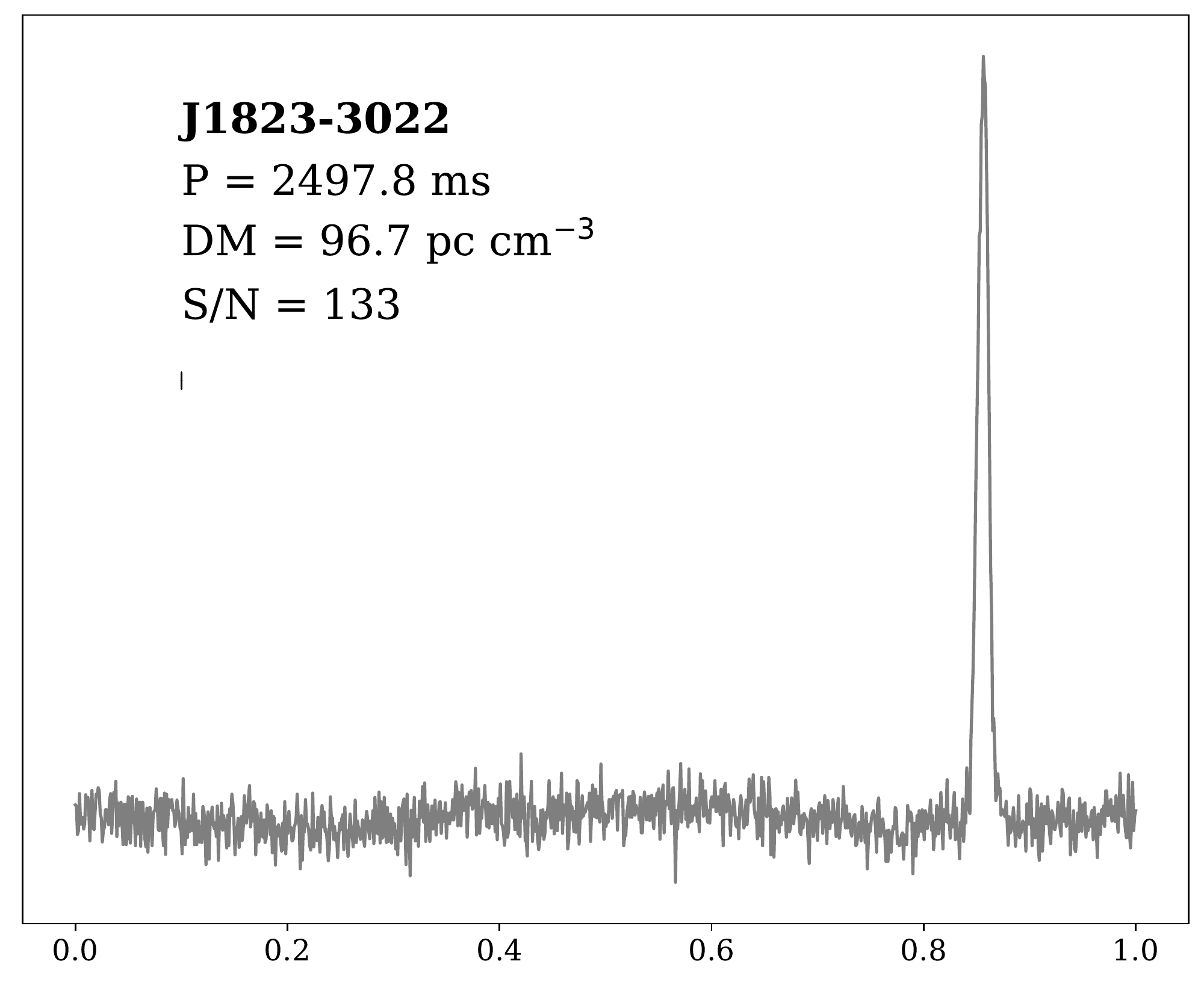}
  	\caption{Profiles of the four pulsars discovered in the TRAPUM observations of NGC 6624. The name, period and DM for each pulsar are reported in each plot. The bar under the DM shows the quadrature sum of the time resolution of 76 $\mu$s of the observations and the DM smearing in the worst affected channel in comparison to the period of the pulsars. The profile of J1823$-$3022 shows variations of the baseline that are caused by broadband RFI.}
  	\label{fig:discovery_profiles}
\end{figure*}

\begin{figure*}
\centering
	\includegraphics[width=\textwidth]{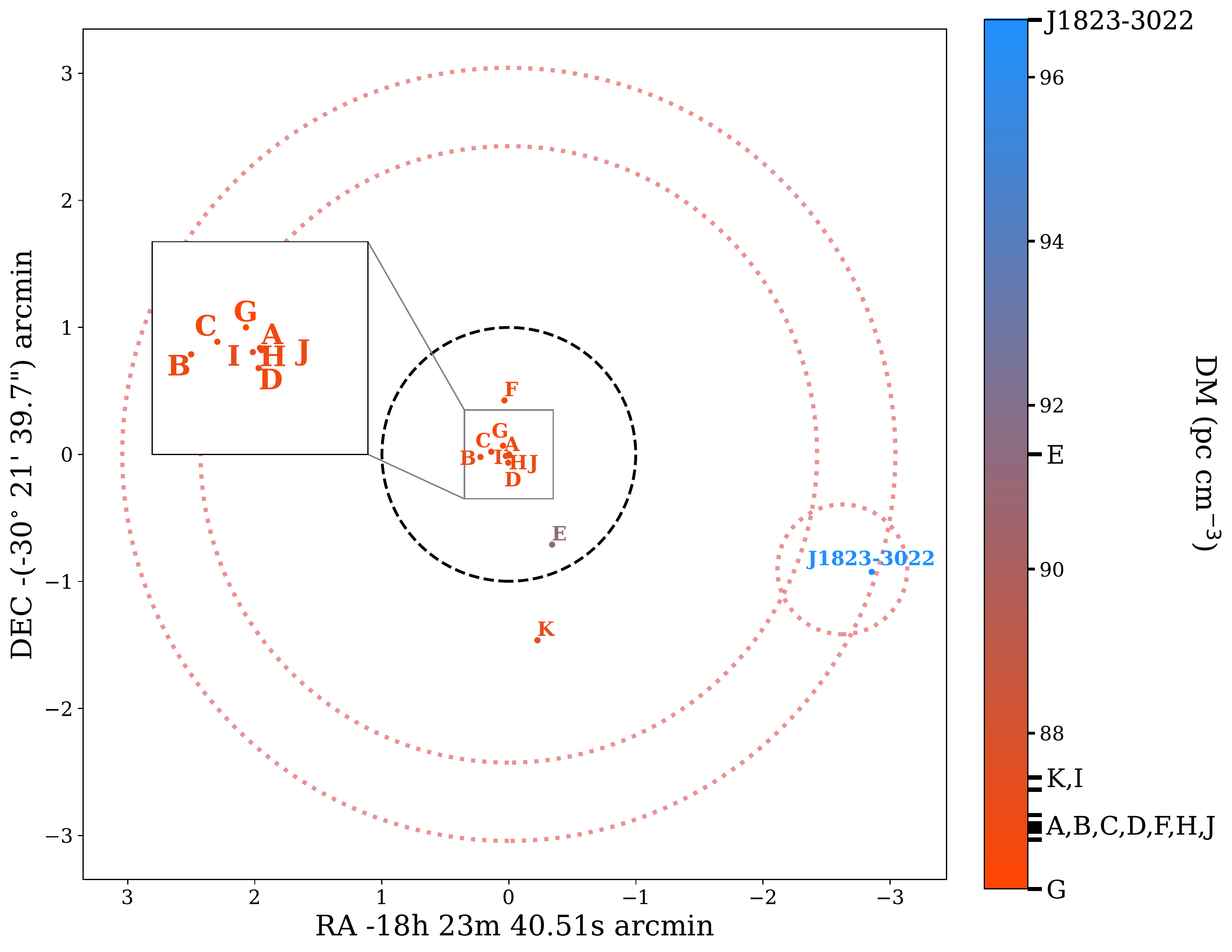}
  	\caption{Map of the cluster NGC 6624 with all the known pulsars. The previously known pulsars are marked with a blue letter while the new discoveries are shown in black. The black dashed circle shows the half-light radius of the cluster while the red dotted curves show the extension of the tiling of the three observations as in Figure \ref{fig:tiling_patterns} and in Figure \ref{fig:tiling_november}. The color of the pulsars represents the value of DM according to the color bar at the right of the plot. Of the new discoveries, pulsar J1823$-$3021I and K have been localised using SeeKAT and J1823-3022 with timing.}
  	\label{fig:positions}
\end{figure*}

Four new pulsars were discovered of which three can safely be considered members of the GC. The profiles of the new discoveries are shown in Fig. \ref{fig:discovery_profiles} while the period, position, values of DM and flux density { are shown in Table \ref{tab:discoveries} and in Fig. \ref{fig:positions}. }

\begin{table*}
\caption{Properties of the newly discovered pulsars in the observations of NGC 6624. We report the barycentric period, DM, position and flux density derived from the radiometer equation. For pulsars I and K we report the position found using the \texttt{SeeKAT} software. For pulsar J the software could not be used and we report the position of the beam at the center of the cluster were the pulsar was found. For pulsar J1823$-$3022 we report the position derived by timing. Numbers in parentheses represent 1-$\sigma$ uncertainties in the last digit.}
\label{tab:discoveries}
\footnotesize
\centering
\renewcommand{\arraystretch}{1.0}
\vskip 0.1cm
\begin{tabular}{lcccllc}
\hline
\hline
\multicolumn{7}{c}{Summary of Discoveries}\\
\hline
Pulsar      & Type        &  \multicolumn{1}{c}{$P$}   & \multicolumn{1}{c}{DM}        &  \multicolumn{1}{c}{$\alpha_{\rm J2000}$}  & \multicolumn{1}{c}{$\delta_{\rm J2000}$} & S$_{1300}$  \\
name        &             & \multicolumn{1}{c}{(ms)}   & \multicolumn{1}{c}{(pc cm$^{-3}$)}  &  & &($\mu$Jy)\\
\hline
NGC 6624I   &  Isolated   & 4.319   & 87.31(4)  & 18$\h$23$\m$40$\s$.6(2) & $-30\degr21\arcmin40\arcsec(3)$ & 11.7 \\ 
NGC 6624J   &  Isolated   & 20.899   & 86.82(7)  & 18$\h$23$\m$40$\s$.5(6) & $-30\degr21\arcmin39\arcsec$(14) & 7.7 \\ 
NGC 6624K   &  Isolated   & 2.768   & 87.50(7)  & 18$\h$23$\m$39$\s$.4(3) & $-30\degr23\arcmin07\arcsec(5)$ & 19.7 \\ 
J1823$-$3022   &  Isolated   & 2497.8   & 96.7(2)  & 18$\h$23$\m$27$\s$.27(1) & $-30\degr22\arcmin35\arcsec(2)$ & 41 \\ 
\hline
\end{tabular}
\end{table*}

\subsection{J1823$-$3021I}

PSR J1823$-$3021I is an isolated MSP with a rotational period of 4.319 ms and DM of 87.31 pc cm$^{-3}$. The profile has a single component with { equivalent width of 0.61 ms.} It was detected by the pipeline in the first observation in the central beam and in two neighbouring beams south of the center. After folding the second observation at the discovery period and DM, the pulsar showed up in the central beam.

The position near the center of the cluster and the value of the DM which is compatible with the other pulsars suggest that this new discovery is associated with the cluster.

\subsection{J1823$-$3021J}

PSR J1823$-$3021J is another isolated pulsar. It has a rotational period of 20.899 ms, DM of 86.82 pc cm$^{-3}$ and was found in the central beam of the second observation. The profile has a single component with { equivalent width of 1.96 ms.} It was confirmed after folding the central beam of the first observation at the discovery period and DM.

The DM is compatible with the other known pulsars and it is located in the core of NGC 6624, this suggests that this pulsar is a member of the cluster. Its rotational period is higher than what we would expect for MSPs if the recycling process was allowed to continue uninterrupted. A possible explanation is that the binary was disrupted by a dynamical interaction during the accretion phase. These types of pulsars are known to be present not only in NGC 6624 (B and C), but also in several other GCs with high encounter rate per binary like e.g. M 15, NGC 6440, NGC 6441 \citep{Freire2008} and NGC 6517 \citep{Lynch2011,Pan2021}.

\subsection{J1823$-$3021K}

PSR J1823$-$3021K is the third new isolated pulsar we found in the cluster. It has a rotational period of 2.768 ms and a DM of 87.50 pc cm$^{-3}$. It has a single broad component with { equivalent width of 0.89 ms.} It was found in a beam of the first observation distant $\sim 1.4$ arcminutes from the center and in two neighbouring beams with lower S/N. It was also found by the pipeline in the four beams of the second observation corresponding to the same location. It is the most rapidly spinning known pulsar in this GC.

This pulsar is located at about $\sim 1.4 $ arcminutes from the center at about $\sim 1.4$ times the half-light radius. Given that the DM is close to the rest of the known pulsars and that it is an isolated MSP, it is reasonable to assume that it is a member of NGC 6624.

\subsection{J1823$-$3022}

PSR J1823$-$3022 is an isolated bright pulsar with a period of 2.4978 seconds and a DM of 96.7 pc cm$^{-3}$. It has a single component with { equivalent width of 49 ms.} It was found near the edge of the tiling of the second observation appearing brightest in the beam on the edge of the tiling and, with lower S/N, in 4 surrounding beams. This position was not covered in the tiling of the first observation but the pulsar still appears by folding at the discovery period and DM in one beam of the first observation as shown in Fig. \ref{fig:tiling_patterns}. 
The pulsar was so bright that it was possible to see single pulses coming from the pulsar. We detected 27 single pulses with S/N higher than 7 corresponding to 0.5 percent of all the rotations of the pulsar over the course of 4 hours of observations. 

\begin{figure*}
\centering
	\includegraphics[width=0.6\textwidth]{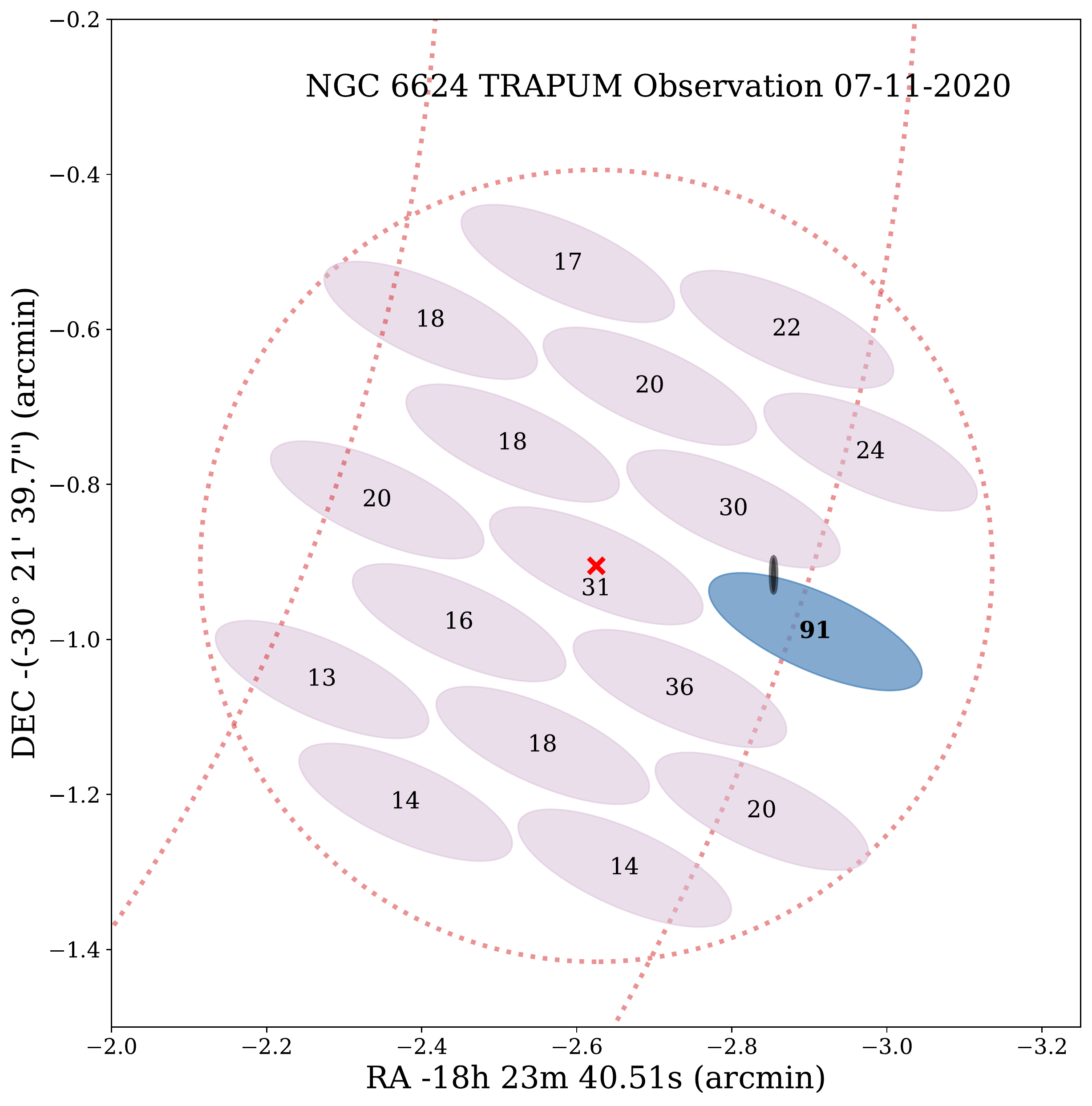}
	\caption{Tiling pattern of the beams of the TRAPUM observations of NGC 6624 performed on the 7th of November 2021 in order to localise J1823$-$3022. The light pink ellipses show the single beams up to 84 percent of the boresight power. The red cross marks the position of the detection with highest S/N of J1823$-$3022 in the previous observations. The number in each beam represents the value of the S/N of the detection. The beam with the highest S/N is highlighted in blue. The red dotted circles are the maximum extent of the size of the tiling in this and the two previous observations plotted for comparison. The black ellipse shows the position of J1823$-$3022 as determined by the timing. The size of ellipse represents the 1$\sigma$ uncertainty on the position.}
  	\label{fig:tiling_november}
\end{figure*}

The pulsar was found with the highest S/N in a beam at the edge of the tiling. It is possible, therefore, that the true position is further away from the cluster center than we measured. In order to localise the source we made an additional observation on the 7th of November 2021 using only 38 antennas in L-band and forming only 17 beams with an average overlap at the central frequency of $\sim$ 0.7 around the position of the beam with the highest S/N as shown in Fig. \ref{fig:tiling_november}. The pulsar was detected in all of the beams of this observation but also in this case the beam with the highest S/N was the furthest from the centre along the edge of the tiling.

Using the detected period and provisional position, we folded archival observations of the cluster made over the years { with MeerKAT, the Parkes radio telescope} and the Green Bank Telescope (GBT). Previous MeerKAT observations carried out with the PTUSE backend had a tied array beam of just $\sim 0.5$ arcminutes while the pulsar appears to be located at $\sim 3$ arcminutes. Despite this, the pulsar was detected in an observation made at UHF-band (550-1050 MHz) where the beam size is significantly larger. We were able to detect the pulsar { in an observation at Parkes from 1999 using the Analog Filterbank in the frequency range 1240-1497 MHz and in a recent one from 2019 using the Pulsar Digital Filterbank 4 in the same frequency range. We were also able to detect the pulsar in observations taken around 2009 at the GBT with the GUPPI backend in the frequency range 1600-2400 MHz and in a set of observations taken around 2020 using the VEGAS backend both in the frequency range 1100-1900 MHz and in the frequency range 1600-2400 MHz. This was possible because the beams of Parkes and GBT in the sky are large enough to cover the position of J1823$-$3022.} The pulsar was not detected sooner in these observations likely because of the bright broadband RFI that appear at these long periods. These RFI can also be seen in the TRAPUM observations and cause the variations in the baseline of the profile shown in Fig. \ref{fig:discovery_profiles}. 

\begin{figure}
\centering
	\includegraphics[width=0.48\textwidth]{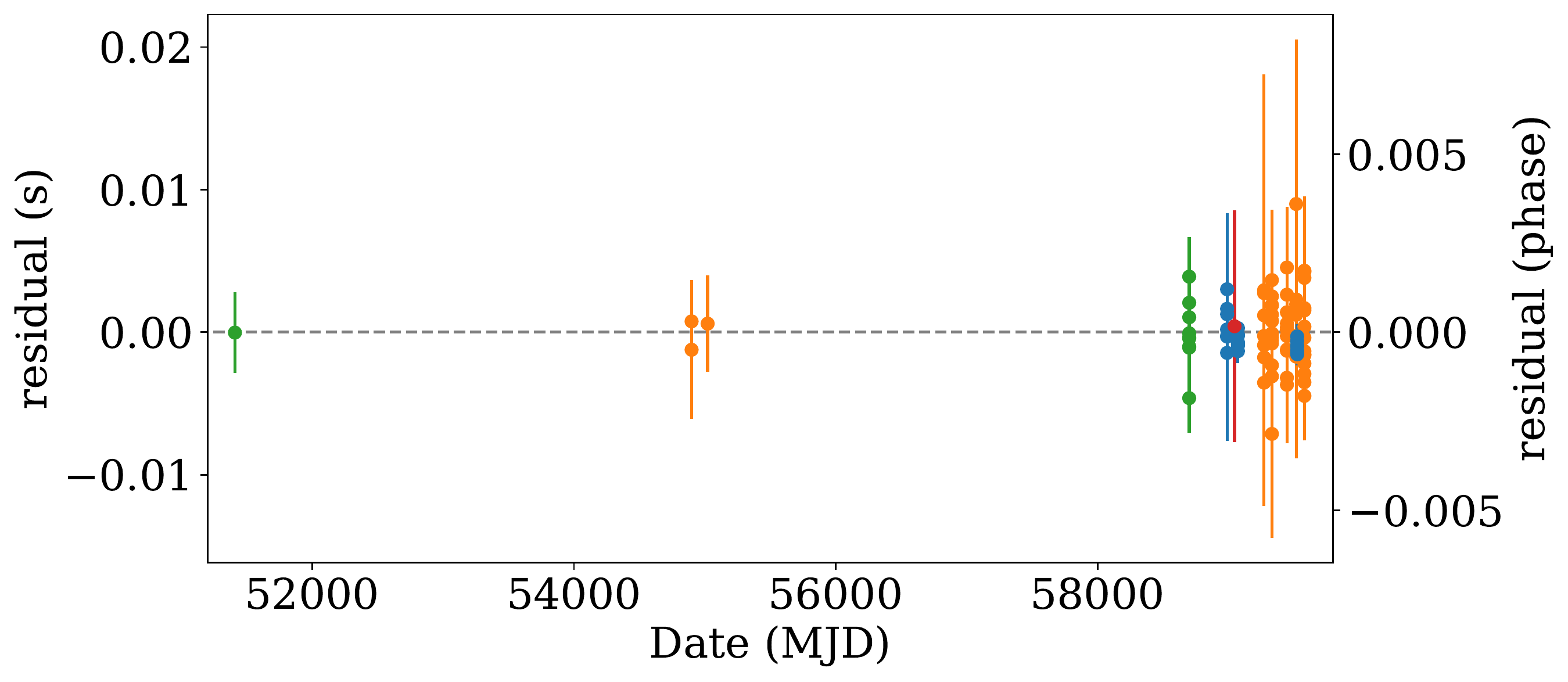}
	\caption{Timing residuals of J1823$-$3022 as a a function of time. The points in blue represent the ToAs obtained from TRAPUM observations at MeerKAT, the red point is from a PTUSE observation at MeerKAT, the orange points are from GBT observations and the green points from observations at Parkes.}
  	\label{fig:residuals_J1823-3022}
\end{figure}

Thanks to these detections, we were able to determine a phase-connected timing solution for J1823$-$3022 that spans over two decades from 1999 to 2022. The timing solution is shown in Table \ref{tab:timing_fitresults} and the residuals of the ToAs are shown in Fig. \ref{fig:residuals_J1823-3022}. The position found coincides with the beam with highest S/N in the observation of the 7th of November 2021 as shown by the black ellipse in Fig. \ref{fig:tiling_november}. 
\begin{table*}
\setlength\tabcolsep{10pt}
    \caption{Timing solution for J1823$-$3022 as derived from the timing of the observed ToA with \texttt{TEMPO2}. The time units are TDB, the adopted terrestrial time standard is TT(TAI) and the Solar System ephemeris used is JPL DE421 \citep{Folkner2009}. Numbers in parentheses represent 1-$\sigma$ uncertainties in the last digit.}
    \centering
     \label{tab:timing_fitresults}
    \renewcommand{\arraystretch}{1.0}
     \begin{tabular}{l c} 
     \hline
      \hline
Pulsar  &   J1823$-$3022    \\
\hline
R.A. (J2000) \dotfill   &  18:23:27.27(1)    \\
DEC. (J2000) \dotfill   & $-$30:22:35(1)   \\
Spin Frequency, $f$ (s$^{-1}$)  \dotfill 	& 0.400354407801(6) \\
1st Spin Frequency derivative, $\dot{f}$ (Hz\,s$^{-1}$) \dotfill & $-$1.3021(2)$\times 10^{-16}$\\
Reference Epoch (MJD)  \dotfill & 59072.811\\
Start of Timing Data (MJD) \dotfill & 51409.581 \\
End of Timing Data (MJD) \dotfill 	& 59580.786 \\
Dispersion Measure (pc\,cm$^{-3}$) \dotfill	& 96.7(2)\\
Number of ToAs \dotfill     & 86\\
Weighted rms residual ($\mu$s) \dotfill  &  2319 \\
$S_{1300}$ (mJy) \dotfill &  0.039\\
\hline
\multicolumn{2}{c}{Derived Parameters}  \\
\hline
Spin Period, $P$ (s)   \dotfill & 2.49778691208(4)\\ 
1st Spin Period derivative, $\dot{P}$ (s\,s$^{-1}$)  \dotfill &  8.123(1)$\times 10^{-16}$ \\
Surface Magnetic Field, $B_0$, (G)  \dotfill & 1.44 $\times 10^{12}$ \\
Characteristic Age, $\tau_{\rm c}$ (Myr) \dotfill & 48.75 \\
    \hline
      \hline
     \end{tabular}
\end{table*}

The position of the pulsar derived by timing places it at $2.99 $ arcminutes from the center or $\sim 3$ times the half-light radius of the cluster. This is almost twice the half-mass radius of $\sim 1.58$ arcminutes \citep{Baumgardt2018}, but still inside the estimated tidal radius of the cluster, corresponding to $\sim 8.33$ arcminutes\footnote{calculated using eq. 8 of \citep{Webb2013} using data from \citep{Baumgardt2019}. For details see: \url{ https://people.smp.uq.edu.au/HolgerBaumgardt/globular/parameter.html}}. 
The presence of pulsars so far from the center of a GC is rare but a few exceptions are known in other core collapsed clusters like B1718-19A \citep{Lyne1993}, located $\sim 2.6$ half-mass radii from the center, J1911-5958A \citep{Damico2001}, distant $\sim 1.4$ half-mass radii from the center, and J1801-0857D, located $\sim 1.4$ \citep{Lynch2011} half-mass radii from the center.

The DM is $\sim 10$ pc cm$^{-3}$ higher than the average DM value than for the pulsars located in the center of the GC and significantly higher than the rest of the known pulsars.
It is outside the range of DM initially searched but it was still detected thanks to its slow period. The incorrect DM causes a delay in the arrival times of the pulse at different frequencies that is in the order of 5 ms for every unit of DM in L-band. This is smaller than the pulse width of the new discovery and therefore does not impact the detectability. Such a high difference in DM has only been seen in two GCs, Terzan 5 and NGC 6517 but in both cases the central value of DM is more than twice that of NGC 6624.

The DM distribution of the known pulsars in the GC is shown in Figure \ref{fig:positions}. Most of the pulsars have DM within 86 and 88 pc cm$^{-3}$ while J1823$-$3021E has a DM significantly higher than the rest at 91.4 pc cm$^{-3}$. J1823$-$3021E is certainly associated with the cluster given that it is an isolated MSP and is located within the half-light radius. We can conclude that the intervening medium along the line of sight can change the DM of the order of 4 pc cm$^{-3}$ over a distance of $\sim$ 1 arcminute. This intervening medium can either be located inside the cluster or in the Galaxy along the line of sight. Applying this argument to J1823$-$3022 we can conclude the DM excess of $\sim 10$ pc cm$^{-3}$ over 3 arcminutes could be explained by foreground material and that the pulsar could be part of the cluster. 
Alternatively we can look at a Galactic model of electron density distribution to estimate the distance of this pulsar based on the value of DM. { Using the  models NE2001 from \cite{Cordes2002} and YMW16 from \cite{Yao2017} we find that the cluster should be located at 2.4 kpc according to NE2001 (3.1 kpc according to YMW16) and the pulsar J1823$-$3022 at 2.6 kpc (3.8 kpc). This would imply that J1823$-$3022 is 0.2 kpc (0.7 kpc) further than the center of the cluster. However, the distance to the cluster estimated this way is a factor of 3 (2) smaller than the distance of the cluster of 8.0 kpc measured from spectroscopy and kinematics  \citep{Baumgardt2021}. This implies that the electron density models largely underestimate the distances along this line of sight and are therefore not reliable regarding the position of J1823$-$3022.}

The period of 2.497 s is very long for a GC pulsar and more typical for Galactic disk pulsars. The slowest pulsar known to be associated with a GC is B1718-19A with a period of 1.002 seconds \citep{Lyne1993}. It must be stated, however, that the GC NGC 6624 already has two other long period pulsars suggesting that apparently young pulsars are able to survive in the cluster.

\begin{figure*}
\centering
	\includegraphics[width=0.6\textwidth]{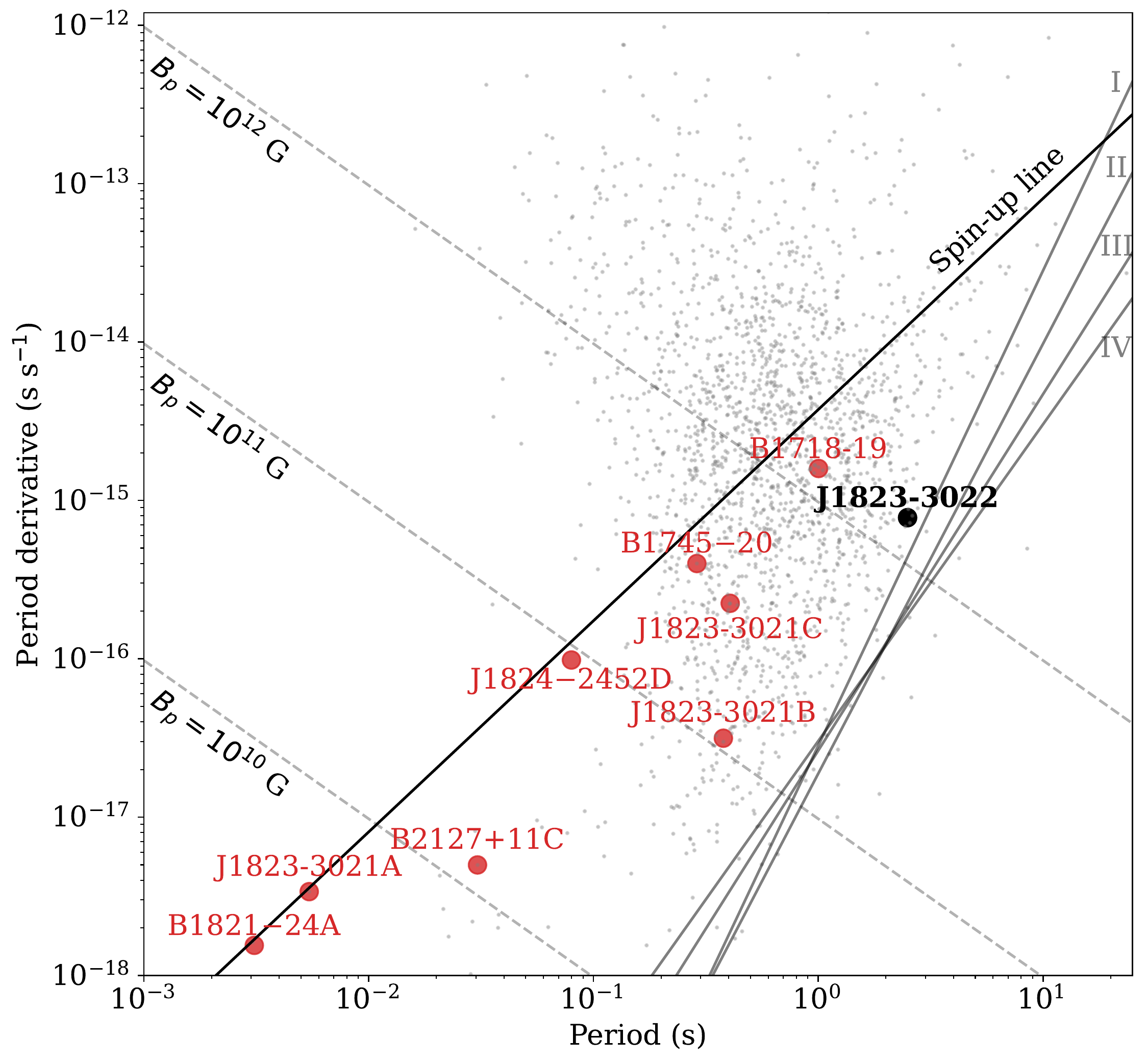}
	\caption{$P$-$\dot{{P}}$ diagram showing the measured period derivative of J1823$-$3022 (black). The plot shows lines of constant magnetic field, the spin-up line as adjusted by \protect\cite{Verbunt2014} to fit B1821$-$24A and J1823$-$3021A, and four models of the death line marked from I to IV following \protect\cite{Cruces2021}. The pulsars associated with GCs with the highest magnetic fields are plotted in red for comparison, for these the effect of the cluster acceleration on the measured $\dot{{P}}$ is significantly smaller than the intrinsic $\dot{{P}}$. { These pulsars were likely old, unrecycled NSs within the clusters that have recently (within the last 2\% of the age of the cluster, as inferred from their characteristic ages) gone through a recycling episode, hence their location below the spin-up line. This was then interrupted by further stellar interactions, otherwise they would have become ``normal'' MSPs in the lower left of the diagram; this is consistent with the fact that we only see these systems in GCs with very high interactions rates per binary. The position of J1823$-$3022 in this diagram is consistent with this hypothesis, but also with it being an unassociated, normal pulsar in the background (grey dots); these} represent all the pulsars present in the PSRCAT catalogue \citep[version 1.66]{Manchester2005}. Note the presence of at least three, possibly four systems associated with NGC 6624 in this plot.}
  	\label{fig:p-pdot_J1823$-$3022}
\end{figure*}

The period derivative measured through timing of J1823$-$3022 is $8.123 \pm 0.001 \, \times 10^{-16}$ s s$^{-1}$. This corresponds to a surface magnetic field of $1.44 \, \times 10^{12}$ G and a characteristic age of $\sim 50$ Myr. The corresponding position in the $P$-$\dot{{P}}$ diagram is shown in Fig. \ref{fig:p-pdot_J1823$-$3022}. If the pulsar is associated with the GC it would be the strongest magnetic field measured in a GC pulsar but still close to that of B1718-19A that has a magnetic field of $1.26 \times 10^{12}$ G \citep{Lyne1993}. The small characteristic age suggests that J1823$-$3022 is either very young or the recycling process stopped recently. If the pulsar is part of the cluster and the recycling process was stopped by a dynamical encounter that also displaced the pulsar, it would also explain the large positional offset as it would not have had time to sink back in the core. Similarly, if the pulsar was born recently after an accretion-induced collapse or an electron capture supernova, it could have been displaced by a kick.

If the pulsar has gone through a recycling episode in the recent past, it should be located below the spin-up line in the $P$-$\dot{{P}}$ diagram, \citep{Verbunt2014}. This line corresponds to the shortest period that can be reached during the recycling process by accretion at the Eddington limit on the pulsar (see \citealt{Pringle1972}):

\begin{equation}\label{eq:spin-up}
{P_{\rm min}}\simeq 1.3~ {\rm s} \times\left(\frac{\rm B}{10^{12}\, \rm G}\right)^{6/7} \simeq 1.6 ~ {\rm s} \times (10^{15} \dot{{P}})^{3/4}
\end{equation}
where we used ${\rm (B/10^{12})}^2 \simeq 10^{15} { P} \dot{{P}}$. Considering the uncertainties in the models behind equation \ref{eq:spin-up}, an alternative spin-up line where $\dot{\rm{P}}$ is 7 times higher for the same $P$ has been suggested \citep{Verbunt2014}. All of the pulsars with high magnetic field strengths in GCs lie below such a line in the $P$-$\dot{{P}}$ diagram. This line is shown in Fig. \ref{fig:p-pdot_J1823$-$3022} alongside some of the pulsars associated with GCs with the highest magnetic fields. J1823$-$3022 is well below this line making it possible that is has gone through a recent recycling episode. However, the period derivative of this pulsar is similar to that of young Galactic pulsar with similar periods.
Based on this parameter we cannot say whether it is associated with the cluster or not.

If the pulsar is not associated we can estimate the probability of it being aligned along the same line of sight. The number of known pulsars in an area of 10 degrees in radius surrounding the cluster according to the \texttt{PSRCAT} catalogue\footnote{\url{https://www.atnf.csiro.au/research/pulsar/psrcat/}}(version 1.66, \citealt{Manchester2005}) is 147. The probability of finding one within 3 arcminutes is $\sim 0.003$ corresponding to about $3\sigma$. Of all of them only 9 have a DM within 10 pc cm$^{-3}$ of the central value of NGC 6624. However, it must be noted that the pulsars in the \texttt{PSRCAT} catalogue are significantly brighter than the newly discovered pulsar so this should be considered as a lower limit to the probability.

\section{Localisation}

The small size of the beams in the TRAPUM observations can determine the position of the newly discovered pulsars with a precision of 10-20 arcseconds. If the new discovery has been detected in more than one beam, it is possible to improve on the localisation using the SeeKAT mulitbeam localiser\footnote{\url{https://github.com/BezuidenhoutMC/SeeKAT}} (Bezuidenhout et al. in prep.). This code compares the S/N of the detections of the various beams with the Point Spread Function (PSF) of the beam calculated using the \texttt{Mosaic} software. The PSF will be dependent on the specific antennas used in the array during the observations, the position in the sky and the frequency. We obtain the best improvement on the localisation when we compare detections from different observations with different PSFs. 

To facilitate the localisation of the pulsars close to the centre of the cluster we performed another observation on the 18th of June 2021 in L-band using 60 antennas with an average overlap of $\sim$ 0.85 covering only the central 1 arcminute with an observation time of 3.5 hours. In this observation we were able to detect both J1823$-$3021I and J1823$-$3021J but only the first was detected in more than one beam. Using this observation it was possible to improve the localisation of J1823$-$3021I. The numerous detections of J1823$-$3021K in the two observations also allow a precise determination of its position. The best positions of the new pulsars are shown in Table \ref{tab:discoveries} and in Fig. \ref{fig:positions}.

\section{Discussion}

The TRAPUM pulsar searches in the GC NGC 6624 have resulted in the discovery of four new pulsars. For one of them, J1823$-$3022, the association with the cluster is not certain. The total number of pulsars with confirmed association in the cluster is now 11. 
All of the new pulsars discovered are isolated even though we searched also for binary pulsars. The pulsar population can therefore be divided in to 9 isolated pulsars and only 2 binaries. This result is quite unusual for a typical GC but it is common for core-collapsed cluster with a very high encounter rate per binary \citep{Verbunt2014} like NGC 6624. Another characteristic of GCs with high encounter rate per binary is the presence of mildly recycled pulsars that were disrupted before the spin-up process was completed. The newly discovered pulsar J1823$-$3021J with a period of 20.899 ms belongs in this category. 

\cite{Verbunt2014} also noted that in GCs with high encounter rate per binary a few pulsars tend to be located far from the central regions. Because of dynamical friction, the neutron stars would concentrate in the cores, but close encounters have the effect of displacing them.
If the close encounters are frequent enough, there is a chance of observing a few pulsars while they are displaced before they drift back to the core.
Thanks to the capability of TRAPUM to localize a pulsar in the cluster, \cite{Ridolfi2021} showed that this is true also in NGC 6624. Pulsar J1823$-$3021E has been found to be offset from the center at a distance of roughly one half-light radius. This is further confirmed by the new discovery of J1823$-$3021K being located $\sim 1.4$ times the half-light radius. These two pulsars have likely been involved in dynamical encounters that displaced them in the recent past and did not have time to fall back in the center.

The newly discovered pulsars certainly associated with the GC are either too weak or too offset to be detected by the large number of observations made of the cluster using up to 44 antennas of MeerKAT with the PTUSE machines and we were not able to derive a timing solution for any of them. With further observations with more antennas it will be possible to find a timing solution for these discoveries. The measurement of the acceleration felt by the two new discoveries close to the center, J1823$-$3021I and J1823$-$3021J, will be very helpful in testing the presence of the intermediate mass black hole claimed by \cite{Perera2017}.

Pulsar J1823$-$3022 is an intriguing new discovery. We do not have enough information to conclude whether the pulsar is associated with the GC. The period, period derivative, isolated nature and position far from the center of the cluster are more common features of the Galactic field pulsar population than the GC population. However, it is still inside the projected tidal radius of NGC 6624 and the cluster already contains two apparently young isolated pulsars with relatively high magnetic field strengths. The high period derivative and offset could be explained by a recent dynamical event that stopped the recycling and kicked the pulsar far from the cluster center. Alternatively the pulsar could be born recently via the process of accretion-induced collapse \citep{Lyne1996} or electron capture supernova \citep{Boyles2011} and have received a kick. In order to determine the pulsar's association with the GC, further observations are needed. A measurement of the proper motion would verify if the pulsar's velocity through space is in agreement with the kinematics of the GC. The discovery of more pulsars associated with the cluster at a similar distance or DM of J1823$-$3022 could also help solve the mystery.

The discovery of J1823$-$3021K and J1823$-$3022, both with a very high positional offset from the center, would not have been possible without the multiple beamforming capabilities and tiling patterns made available by TRAPUM. This shows that a significant number of interesting and potentially bright pulsars might still be waiting to be found in the outskirts of well known clusters. Among the GC surveys currently in progress at the various telescopes around the world, TRAPUM has the best chances of finding and localizing these pulsars.

\section*{Acknowledgements}

The MeerKAT telescope is operated by the South African Radio Astronomy Observatory, which is a facility of the National Research Foundation, an agency of the Department of Science and Innovation. SARAO acknowledges the ongoing advice and calibration of GPS systems by the National Metrology Institute of South Africa (NMISA) and the time space reference systems department of the Paris Observatory.
PTUSE was developed with support from  the Australian SKA Office and Swinburne University of Technology. MeerTime data is housed on the OzSTAR supercomputer at Swinburne University of Technology. The OzSTAR program receives funding in part from the Astronomy National Collaborative Research Infrastructure Strategy (NCRIS) allocation provided by the Australian Government.
The authors also acknowledge MPIfR funding to contribute to MeerTime infrastructure.
TRAPUM observations used the FBFUSE and APSUSE computing clusters for data acquisition, storage and analysis. These clusters were funded and installed by the Max-Planck-Institut für Radioastronomie and the Max-Planck-Gesellschaft. The National Radio Astronomy Observatory is a facility of the National Science Foundation operated under cooperative agreement by Associated Universities, Inc. The Green Bank Observatory is a facility of the National Science Foundation operated under cooperative agreement by Associated Universities, Inc. 
FA, AR, EDB, DJC, WC, PCCF, TG, MK, PVP and VVK  acknowledge continuing valuable support from the Max-Planck Society.
FA gratefully acknowledges support from ERC Synergy Grant `BlackHoleCam' Grant Agreement Number 610058.
SMR is a CIFAR Fellow and is supported by the NSF Physics Frontiers Center awards 1430284 and 2020265.
This work is supported by the Max-Planck Society as part of the "LEGACY" collaboration on low-frequency gravitational wave astronomy.
AR MBu and AP gratefully acknowledge financial support by the research grant ``iPeska'' (P.I. Andrea Possenti) funded under the INAF national call Prin-SKA/CTA approved with the Presidential Decree 70/2016. AR, MBu and AP also acknowledge support from the Ministero degli Affari Esteri e della Cooperazione Internazionale - Direzione Generale per la Promozione del Sistema Paese - Progetto di Grande Rilevanza ZA18GR02.
BWS acknowledges funding from the European Research Council (ERC) under the European Union’s Horizon 2020 research and innovation programme (grant agreement No. 694745). RPB acknowledges support ERC Starter Grant `Spiders', Grant Agreement Number 715051.

\section*{Data Availability}

The data underlying this article will be shared upon reasonable request to the MeerTime and TRAPUM collaborations.



\bibliographystyle{mnras}
\bibliography{biblio} 

\begin{thebibliography}{}
\makeatletter
\relax
\def\mn@urlcharsother{\let\do\@makeother \do\$\do\&\do\#\do\^\do\_\do\%\do\~}
\def\mn@doi{\begingroup\mn@urlcharsother \@ifnextchar [ {\mn@doi@}
  {\mn@doi@[]}}
\def\mn@doi@[#1]#2{\def\@tempa{#1}\ifx\@tempa\@empty \href
  {http://dx.doi.org/#2} {doi:#2}\else \href {http://dx.doi.org/#2} {#1}\fi
  \endgroup}
\def\mn@eprint#1#2{\mn@eprint@#1:#2::\@nil}
\def\mn@eprint@arXiv#1{\href {http://arxiv.org/abs/#1} {{\tt arXiv:#1}}}
\def\mn@eprint@dblp#1{\href {http://dblp.uni-trier.de/rec/bibtex/#1.xml}
  {dblp:#1}}
\def\mn@eprint@#1:#2:#3:#4\@nil{\def\@tempa {#1}\def\@tempb {#2}\def\@tempc
  {#3}\ifx \@tempc \@empty \let \@tempc \@tempb \let \@tempb \@tempa \fi \ifx
  \@tempb \@empty \def\@tempb {arXiv}\fi \@ifundefined
  {mn@eprint@\@tempb}{\@tempb:\@tempc}{\expandafter \expandafter \csname
  mn@eprint@\@tempb\endcsname \expandafter{\@tempc}}}

\bibitem[\protect\citeauthoryear{{Abbate} et~al.,}{{Abbate}
  et~al.}{2020}]{Abbate2020}
{Abbate} F.,  et~al., 2020, \mn@doi [\mnras] {10.1093/mnras/staa2510}, \href
  {https://ui.adsabs.harvard.edu/abs/2020MNRAS.498..875A} {498, 875}

\bibitem[\protect\citeauthoryear{{Bahramian}, {Heinke}, {Sivakoff}  \&
  {Gladstone}}{{Bahramian} et~al.}{2013}]{Bahramian2013}
{Bahramian} A.,  {Heinke} C.~O.,  {Sivakoff} G.~R.,   {Gladstone} J.~C.,  2013,
  \mn@doi [\apj] {10.1088/0004-637X/766/2/136}, \href
  {https://ui.adsabs.harvard.edu/abs/2013ApJ...766..136B} {766, 136}

\bibitem[\protect\citeauthoryear{{Bailes} et~al.,}{{Bailes}
  et~al.}{2016}]{Bailes2016}
{Bailes} M.,  et~al., 2016, in MeerKAT Science: On the Pathway to the SKA.
  p.~11 (\mn@eprint {arXiv} {1803.07424})

\bibitem[\protect\citeauthoryear{{Bailes} et~al.,}{{Bailes}
  et~al.}{2020}]{Bailes2020}
{Bailes} M.,  et~al., 2020, \mn@doi [\pasa] {10.1017/pasa.2020.19}, \href
  {https://ui.adsabs.harvard.edu/abs/2020PASA...37...28B} {37, e028}

\bibitem[\protect\citeauthoryear{{Baumgardt} \& {Hilker}}{{Baumgardt} \&
  {Hilker}}{2018}]{Baumgardt2018}
{Baumgardt} H.,  {Hilker} M.,  2018, \mn@doi [\mnras] {10.1093/mnras/sty1057},
  \href {https://ui.adsabs.harvard.edu/abs/2018MNRAS.478.1520B} {478, 1520}

\bibitem[\protect\citeauthoryear{{Baumgardt} \& {Vasiliev}}{{Baumgardt} \&
  {Vasiliev}}{2021}]{Baumgardt2021}
{Baumgardt} H.,  {Vasiliev} E.,  2021, \mn@doi [\mnras]
  {10.1093/mnras/stab1474}, \href
  {https://ui.adsabs.harvard.edu/abs/2021MNRAS.505.5957B} {505, 5957}

\bibitem[\protect\citeauthoryear{{Baumgardt}, {Hilker}, {Sollima}  \&
  {Bellini}}{{Baumgardt} et~al.}{2019}]{Baumgardt2019}
{Baumgardt} H.,  {Hilker} M.,  {Sollima} A.,   {Bellini} A.,  2019, \mn@doi
  [\mnras] {10.1093/mnras/sty2997}, \href
  {https://ui.adsabs.harvard.edu/abs/2019MNRAS.482.5138B} {482, 5138}

\bibitem[\protect\citeauthoryear{{Biggs}, {Bailes}, {Lyne}, {Goss}  \&
  {Fruchter}}{{Biggs} et~al.}{1994}]{Biggs1994}
{Biggs} J.~D.,  {Bailes} M.,  {Lyne} A.~G.,  {Goss} W.~M.,   {Fruchter} A.~S.,
  1994, \mn@doi [\mnras] {10.1093/mnras/267.1.125}, \href
  {https://ui.adsabs.harvard.edu/abs/1994MNRAS.267..125B} {267, 125}

\bibitem[\protect\citeauthoryear{{Boyles}, {Lorimer}, {Turk}, {Mnatsakanov},
  {Lynch}, {Ransom}, {Freire}  \& {Belczynski}}{{Boyles}
  et~al.}{2011}]{Boyles2011}
{Boyles} J.,  {Lorimer} D.~R.,  {Turk} P.~J.,  {Mnatsakanov} R.,  {Lynch}
  R.~S.,  {Ransom} S.~M.,  {Freire} P.~C.,   {Belczynski} K.,  2011, \mn@doi
  [\apj] {10.1088/0004-637X/742/1/51}, \href
  {https://ui.adsabs.harvard.edu/abs/2011ApJ...742...51B} {742, 51}

\bibitem[\protect\citeauthoryear{{Camilo} et~al.,}{{Camilo}
  et~al.}{2018}]{Camilo2018}
{Camilo} F.,  et~al., 2018, \mn@doi [\apj] {10.3847/1538-4357/aab35a}, \href
  {https://ui.adsabs.harvard.edu/abs/2018ApJ...856..180C} {856, 180}

\bibitem[\protect\citeauthoryear{{Chen}, {Barr}, {Karuppusamy}, {Kramer}  \&
  {Stappers}}{{Chen} et~al.}{2021}]{Chen2021}
{Chen} W.,  {Barr} E.,  {Karuppusamy} R.,  {Kramer} M.,   {Stappers} B.,  2021,
  arXiv e-prints, \href {https://ui.adsabs.harvard.edu/abs/2021arXiv211001667C}
  {p. arXiv:2110.01667}

\bibitem[\protect\citeauthoryear{{Cordes} \& {Lazio}}{{Cordes} \&
  {Lazio}}{2002}]{Cordes2002}
{Cordes} J.~M.,  {Lazio} T.~J.~W.,  2002, arXiv e-prints, \href
  {https://ui.adsabs.harvard.edu/abs/2002astro.ph..7156C} {pp
  astro--ph/0207156}

\bibitem[\protect\citeauthoryear{{Cruces} et~al.,}{{Cruces}
  et~al.}{2021}]{Cruces2021}
{Cruces} M.,  et~al., 2021, \mn@doi [\mnras] {10.1093/mnras/stab2540}, \href
  {https://ui.adsabs.harvard.edu/abs/2021MNRAS.508..300C} {508, 300}

\bibitem[\protect\citeauthoryear{{D'Amico}, {Lyne}, {Manchester}, {Possenti}
  \& {Camilo}}{{D'Amico} et~al.}{2001}]{Damico2001}
{D'Amico} N.,  {Lyne} A.~G.,  {Manchester} R.~N.,  {Possenti} A.,   {Camilo}
  F.,  2001, \mn@doi [\apjl] {10.1086/319096}, \href
  {https://ui.adsabs.harvard.edu/abs/2001ApJ...548L.171D} {548, L171}

\bibitem[\protect\citeauthoryear{{Folkner}, {Williams}  \& {Boggs}}{{Folkner}
  et~al.}{2009}]{Folkner2009}
{Folkner} W.~M.,  {Williams} J.~G.,   {Boggs} D.~H.,  2009, Interplanetary
  Network Progress Report, \href
  {https://ui.adsabs.harvard.edu/abs/2009IPNPR.178C...1F} {42-178, 1}

\bibitem[\protect\citeauthoryear{{Freire}, {Ransom}, {B{\'e}gin}, {Stairs},
  {Hessels}, {Frey}  \& {Camilo}}{{Freire} et~al.}{2008}]{Freire2008}
{Freire} P. C.~C.,  {Ransom} S.~M.,  {B{\'e}gin} S.,  {Stairs} I.~H.,
  {Hessels} J. W.~T.,  {Frey} L.~H.,   {Camilo} F.,  2008, \mn@doi [\apj]
  {10.1086/526338}, \href
  {https://ui.adsabs.harvard.edu/abs/2008ApJ...675..670F} {675, 670}

\bibitem[\protect\citeauthoryear{{Freire} et~al.,}{{Freire}
  et~al.}{2011}]{Freire2011}
{Freire} P.~C.~C.,  et~al., 2011, \mn@doi [Science] {10.1126/science.1207141},
  \href {https://ui.adsabs.harvard.edu/abs/2011Sci...334.1107F} {334, 1107}

\bibitem[\protect\citeauthoryear{{Hobbs}, {Edwards}  \& {Manchester}}{{Hobbs}
  et~al.}{2006}]{Hobbs2006}
{Hobbs} G.~B.,  {Edwards} R.~T.,   {Manchester} R.~N.,  2006, \mn@doi [\mnras]
  {10.1111/j.1365-2966.2006.10302.x}, \href
  {https://ui.adsabs.harvard.edu/abs/2006MNRAS.369..655H} {369, 655}

\bibitem[\protect\citeauthoryear{{Hotan}, {van Straten}  \&
  {Manchester}}{{Hotan} et~al.}{2004}]{Hotan2004}
{Hotan} A.~W.,  {van Straten} W.,   {Manchester} R.~N.,  2004, \mn@doi [\pasa]
  {10.1071/AS04022}, \href
  {https://ui.adsabs.harvard.edu/abs/2004PASA...21..302H} {21, 302}

\bibitem[\protect\citeauthoryear{{Jonas} \& {MeerKAT Team}}{{Jonas} \& {MeerKAT
  Team}}{2016}]{Jonas2016}
{Jonas} J.,  {MeerKAT Team} 2016, in MeerKAT Science: On the Pathway to the
  SKA. p.~1

\bibitem[\protect\citeauthoryear{{Lorimer} \& {Kramer}}{{Lorimer} \&
  {Kramer}}{2012}]{Handbook2012}
{Lorimer} D.~R.,  {Kramer} M.,  2012, {Handbook of Pulsar Astronomy}.
{Cambridge University Press}

\bibitem[\protect\citeauthoryear{{Lynch}, {Ransom}, {Freire}  \&
  {Stairs}}{{Lynch} et~al.}{2011}]{Lynch2011}
{Lynch} R.~S.,  {Ransom} S.~M.,  {Freire} P. C.~C.,   {Stairs} I.~H.,  2011,
  \mn@doi [\apj] {10.1088/0004-637X/734/2/89}, \href
  {https://ui.adsabs.harvard.edu/abs/2011ApJ...734...89L} {734, 89}

\bibitem[\protect\citeauthoryear{{Lynch}, {Freire}, {Ransom}  \&
  {Jacoby}}{{Lynch} et~al.}{2012}]{Lynch2012}
{Lynch} R.~S.,  {Freire} P. C.~C.,  {Ransom} S.~M.,   {Jacoby} B.~A.,  2012,
  \mn@doi [\apj] {10.1088/0004-637X/745/2/109}, \href
  {https://ui.adsabs.harvard.edu/abs/2012ApJ...745..109L} {745, 109}

\bibitem[\protect\citeauthoryear{{Lyne}, {Biggs}, {Harrison}  \&
  {Bailes}}{{Lyne} et~al.}{1993}]{Lyne1993}
{Lyne} A.~G.,  {Biggs} J.~D.,  {Harrison} P.~A.,   {Bailes} M.,  1993, \mn@doi
  [\nat] {10.1038/361047a0}, \href
  {https://ui.adsabs.harvard.edu/abs/1993Natur.361...47L} {361, 47}

\bibitem[\protect\citeauthoryear{{Lyne}, {Manchester}  \& {D'Amico}}{{Lyne}
  et~al.}{1996}]{Lyne1996}
{Lyne} A.~G.,  {Manchester} R.~N.,   {D'Amico} N.,  1996, \mn@doi [\apjl]
  {10.1086/309972}, \href
  {https://ui.adsabs.harvard.edu/abs/1996ApJ...460L..41L} {460, L41}

\bibitem[\protect\citeauthoryear{{Manchester}, {Hobbs}, {Teoh}  \&
  {Hobbs}}{{Manchester} et~al.}{2005}]{Manchester2005}
{Manchester} R.~N.,  {Hobbs} G.~B.,  {Teoh} A.,   {Hobbs} M.,  2005, VizieR
  Online Data Catalog, \href
  {https://ui.adsabs.harvard.edu/abs/2005yCat.7245....0M} {p. VII/245}

\bibitem[\protect\citeauthoryear{{Morello}, {Rajwade}  \& {Stappers}}{{Morello}
  et~al.}{2021}]{Morello2021}
{Morello} V.,  {Rajwade} K.~M.,   {Stappers} B.~W.,  2021, \mn@doi [\mnras]
  {10.1093/mnras/stab3493}, \href
  {https://ui.adsabs.harvard.edu/abs/2021MNRAS.tmp.3198M} {}

\bibitem[\protect\citeauthoryear{{Noyola} \& {Gebhardt}}{{Noyola} \&
  {Gebhardt}}{2006}]{Noyola2006}
{Noyola} E.,  {Gebhardt} K.,  2006, \mn@doi [\aj] {10.1086/505390}, \href
  {https://ui.adsabs.harvard.edu/abs/2006AJ....132..447N} {132, 447}

\bibitem[\protect\citeauthoryear{{Pan} et~al.,}{{Pan} et~al.}{2021}]{Pan2021}
{Pan} Z.,  et~al., 2021, \mn@doi [\apjl] {10.3847/2041-8213/ac0bbd}, \href
  {https://ui.adsabs.harvard.edu/abs/2021ApJ...915L..28P} {915, L28}

\bibitem[\protect\citeauthoryear{{Perera} et~al.,}{{Perera}
  et~al.}{2017a}]{Perera2017}
{Perera} B.~B.~P.,  et~al., 2017a, \mn@doi [\mnras] {10.1093/mnras/stx501},
  \href {https://ui.adsabs.harvard.edu/abs/2017MNRAS.468.2114P} {468, 2114}

\bibitem[\protect\citeauthoryear{{Perera} et~al.,}{{Perera}
  et~al.}{2017b}]{Perera2017b}
{Perera} B.~B.~P.,  et~al., 2017b, \mn@doi [\mnras] {10.1093/mnras/stx1236},
  \href {https://ui.adsabs.harvard.edu/abs/2017MNRAS.471.1258P} {471, 1258}

\bibitem[\protect\citeauthoryear{{Pringle} \& {Rees}}{{Pringle} \&
  {Rees}}{1972}]{Pringle1972}
{Pringle} J.~E.,  {Rees} M.~J.,  1972, \aap, \href
  {https://ui.adsabs.harvard.edu/abs/1972A&A....21....1P} {21, 1}

\bibitem[\protect\citeauthoryear{{Ransom}, {Eikenberry}  \&
  {Middleditch}}{{Ransom} et~al.}{2002}]{Ransom2002}
{Ransom} S.~M.,  {Eikenberry} S.~S.,   {Middleditch} J.,  2002, \mn@doi [\aj]
  {10.1086/342285}, \href
  {https://ui.adsabs.harvard.edu/abs/2002AJ....124.1788R} {124, 1788}

\bibitem[\protect\citeauthoryear{{Ransom}, {Cordes}  \& {Eikenberry}}{{Ransom}
  et~al.}{2003}]{Ransom2003}
{Ransom} S.~M.,  {Cordes} J.~M.,   {Eikenberry} S.~S.,  2003, \mn@doi [\apj]
  {10.1086/374806}, \href
  {https://ui.adsabs.harvard.edu/abs/2003ApJ...589..911R} {589, 911}

\bibitem[\protect\citeauthoryear{{Ridolfi} et~al.,}{{Ridolfi}
  et~al.}{2021}]{Ridolfi2021}
{Ridolfi} A.,  et~al., 2021, \mn@doi [\mnras] {10.1093/mnras/stab790}, \href
  {https://ui.adsabs.harvard.edu/abs/2021MNRAS.504.1407R} {504, 1407}

\bibitem[\protect\citeauthoryear{{Sosin} \& {King}}{{Sosin} \&
  {King}}{1995}]{Sosin1995}
{Sosin} C.,  {King} I.~R.,  1995, \mn@doi [\aj] {10.1086/117307}, \href
  {https://ui.adsabs.harvard.edu/abs/1995AJ....109..639S} {109, 639}

\bibitem[\protect\citeauthoryear{{Stappers} \& {Kramer}}{{Stappers} \&
  {Kramer}}{2016}]{Stappers2016}
{Stappers} B.,  {Kramer} M.,  2016, in MeerKAT Science: On the Pathway to the
  SKA. p.~9

\bibitem[\protect\citeauthoryear{{Verbunt} \& {Freire}}{{Verbunt} \&
  {Freire}}{2014}]{Verbunt2014}
{Verbunt} F.,  {Freire} P. C.~C.,  2014, \mn@doi [\aap]
  {10.1051/0004-6361/201321177}, \href
  {https://ui.adsabs.harvard.edu/abs/2014A&A...561A..11V} {561, A11}

\bibitem[\protect\citeauthoryear{{Webb}, {Harris}, {Sills}  \& {Hurley}}{{Webb}
  et~al.}{2013}]{Webb2013}
{Webb} J.~J.,  {Harris} W.~E.,  {Sills} A.,   {Hurley} J.~R.,  2013, \mn@doi
  [\apj] {10.1088/0004-637X/764/2/124}, \href
  {https://ui.adsabs.harvard.edu/abs/2013ApJ...764..124W} {764, 124}

\bibitem[\protect\citeauthoryear{{Yao}, {Manchester}  \& {Wang}}{{Yao}
  et~al.}{2017}]{Yao2017}
{Yao} J.~M.,  {Manchester} R.~N.,   {Wang} N.,  2017, \mn@doi [\apj]
  {10.3847/1538-4357/835/1/29}, \href
  {https://ui.adsabs.harvard.edu/abs/2017ApJ...835...29Y} {835, 29}

\bibitem[\protect\citeauthoryear{{van Straten} \& {Bailes}}{{van Straten} \&
  {Bailes}}{2011}]{vanStraten2011}
{van Straten} W.,  {Bailes} M.,  2011, \mn@doi [\pasa] {10.1071/AS10021}, \href
  {https://ui.adsabs.harvard.edu/abs/2011PASA...28....1V} {28, 1}

\bibitem[\protect\citeauthoryear{{van Straten}, {Demorest}  \& {Oslowski}}{{van
  Straten} et~al.}{2012}]{vanStraten2012}
{van Straten} W.,  {Demorest} P.,   {Oslowski} S.,  2012, Astronomical Research
  and Technology, \href {https://ui.adsabs.harvard.edu/abs/2012AR&T....9..237V}
  {9, 237}

\makeatother
\end{thebibliography}





\bsp	
\label{lastpage}
\end{document}